\documentclass{PoS}
\def\araa{ARA\&A}
\def\apj{ApJ}
\def\apjs{ApJS}
\def\aap{A\&A}
\def\mnras{MNRAS}
\def\nat{Nature}

\usepackage{lineno}

\title{Gamma-ray light curves for the BL Lac Mrk 421 using  HAWC data derived with a new approach.}

\ShortTitle{ZEBRA fitting framework}

\author{\speaker{J. A. Garc\'ia-Gonz\'alez}\\
        Instituto de F\'isica, Universidad Nacional Aut\'{o}noma de M\'{e}xico, Apdo. Postal 70-264, Cd. Universitaria, M\'{e}xico DF 04510.\\
        E-mail: \email{jagarcia@fisica.unam.mx}}

\author{M. M. Gonz\'alez\\
        Instituto de Astronom\'ia, Universidad Nacional Aut\'{o}noma de M\'{e}xico, Apdo. Postal 70-264, Cd. Universitaria, M\'{e}xico DF 04510.\\
        E-mail: \email{magda@astro.unam.mx}}
\author{N. Fraija\\
        Instituto de Astronom\'ia, Universidad Nacional Aut\'{o}noma de M\'{e}xico, Apdo. Postal 70-264, Cd. Universitaria, M\'{e}xico DF 04510.\\
        E-mail: \email{nifraija@astro.unam.mx}}

\author{for the HAWC Collaboration\footnote{for collaboration list see PoS(ICRC2019)1177}}

\abstract{
The HAWC gamma ray observatory is located at the Sierra Negra Volcano in Puebla, Mexico, at an altitude of 4,100 meters. HAWC is a wide field of view array of 300 water Cherenkov detectors that are continuously surveying ~ 2sr of the sky since March 2015. The large collected data sample consisting in more than 3 years makes HAWC an ideal instrument to perform an unbiased monitoring of blazars in the very-high-energy (VHE) emission. This is particular relevant for Mrk 421, one of the closest and brightest blazars in the gamma-ray/X-ray classified as high-synchrotron-peaked BL Lac class. In this work we present light curves for Mrk 421 and the Crab nebula obtained with the first 17 months of data and a new analysis framework. We compare the results with the light curves reported in \cite{2017ApJ...841..100A}. The main advantage of the new framework is the capability to derive fluxes for arbitrary timescales. We show that both, previous and present, methods give consistent results. 
}

\FullConference{36th International Cosmic Ray Conference -ICRC2019-\\
		July 24th - August 1st, 2019\\
		Madison, WI, U.S.A.}

\begin{document}

\section{Introduction\label{sec:intro}}
Blazars are a subclass of radio loud active galactic nuclei (AGNs) characterized by having relativistic jets pointing close to the observer's line of sight~\cite{1979ApJ...232...34B}. They are known to exhibit large variability in all spectral bands \cite{1997ARA&A..35..445U}. Many variability studies in blazars have been performed with the aim of exploring variability both on short and long timescales\cite{1997ARA&A..35..445U,Falomo,2017ApJS..232....7F,2015APh....71....1F}. Blazars are usually classified in BL Lac sources and Flat Spectrum Radio Quasars (FSRQ) depending on the equivalent width of the optical emission lines~\cite{1996MNRAS.281..425M,Beckmann}. 

At a distance of 134.1 Mpc, the BL Lac Mrk\ 421 \cite{2005ApJ...635..173S} is one of the closest and most comprehensively studied source of the high-synchrotron-peaked BL Lac (HBL) class. Mrk\,421 is also one of the brightest and most variable extragalactic gamma-ray sources in very-high energies (VHEs) \cite{1992Natur.358..477P}.  
For instance, in the TeV band this object has exhibited extremely large variability, down to timescales of 15 minutes making it an excellent source to be studied by the High-Altitude Water Cherenkov (HAWC) observatory.  

Previously, the HAWC collaboration presented in ~\cite{2017ApJ...841..100A} light curves (LCs) for three of the main sources being monitored by this observatory: the Crab, Mrk 421 and Mrk 501. The analysis was performed with data in the energy range from $300$ GeV to $100$ TeV and using the first 17 months of HAWC operations. The likelihood method~\cite{2015ICRC...34..948Y} was used to construct daily maps on sidereal days, then a specific spectral model was used to count the photon fluxes per every transit for the three sources.


In this work, the maximum likelihood approach is replaced with a new method using a set of tools arranged in a package named Zenith Band Response Analysis (ZEBRA). It uses Monte Carlo simulations to characterize the detector response as a function of zenith angle. Then, the detector response is convolved with the exposure information to estimate the counts observed from a source during an arbitrary period of time. The main
difference with the method used before is the capability to derive flux measurements on shorter time periods. We perform a comparison between the two methods to show that they provide consistent flux measurements.  Section~\ref{sec:fitfram} describes the main features of the two fitting frameworks available to obtain flux measurements. In Section~\ref{sec:lcs}, LC comparison between the two methods is presented for both, the Crab and Mrk\ 421. Finally in Section~\ref{sec:conc} conclusions are presented. 

\section{Fitting frameworks\label{sec:fitfram}}
\subsection{LiFF}
The maximum likelihood method is implemented in the package named Likelihood Fitting Framework (LiFF) that is described in detail elsewhere\cite{2015ICRC...34..948Y}. This method uses a physics spectral model and a detector response model for a given data set. It uses the log likelihood function to estimate the likelihood ratio test (TS) to compare a null hypothesis with the alternative hypothesis. In the case where the null hypothesis is true the TS looks like a $\chi^{2}$ distribution (with N=1 degrees of freedom) and the $\sqrt{TS}$ can be considered as the statistical significance that the null hypothesis can be rejected with. One limitation of this method is that it computes the detector response as a function of declination, assuming a minimum exposure on a full transit of the source, thus making difficult to obtain a result that requires a time scale of hours of less e.g. intra-night variability studies or hardening of spectra during periods of high activity such as the ones presented in $\gamma$-ray flares in blazars. These type of analyses were previously done with a correction factor~\cite{2017ApJ...841..100A}, but still required exposure during the majority of the transit.

\subsection{ZEBRA}
ZEBRA on the other hand proposes a different approach. In order to solve the problem of calculating the flux in time windows smaller than a full transit, ZEBRA computes the detector response as a function of zenith, which is then convolved with the time each location was exposed to each zenith angle. This includes updating the point spread function, strongly dependent on the zenith angle and whose effect is not solved by an overall correction factor, as previously done.

\section{Light Curves: Crab and Mrk 421\label{sec:lcs}}

\begin{figure*}
\begin{center}
\includegraphics[width=0.75\linewidth]{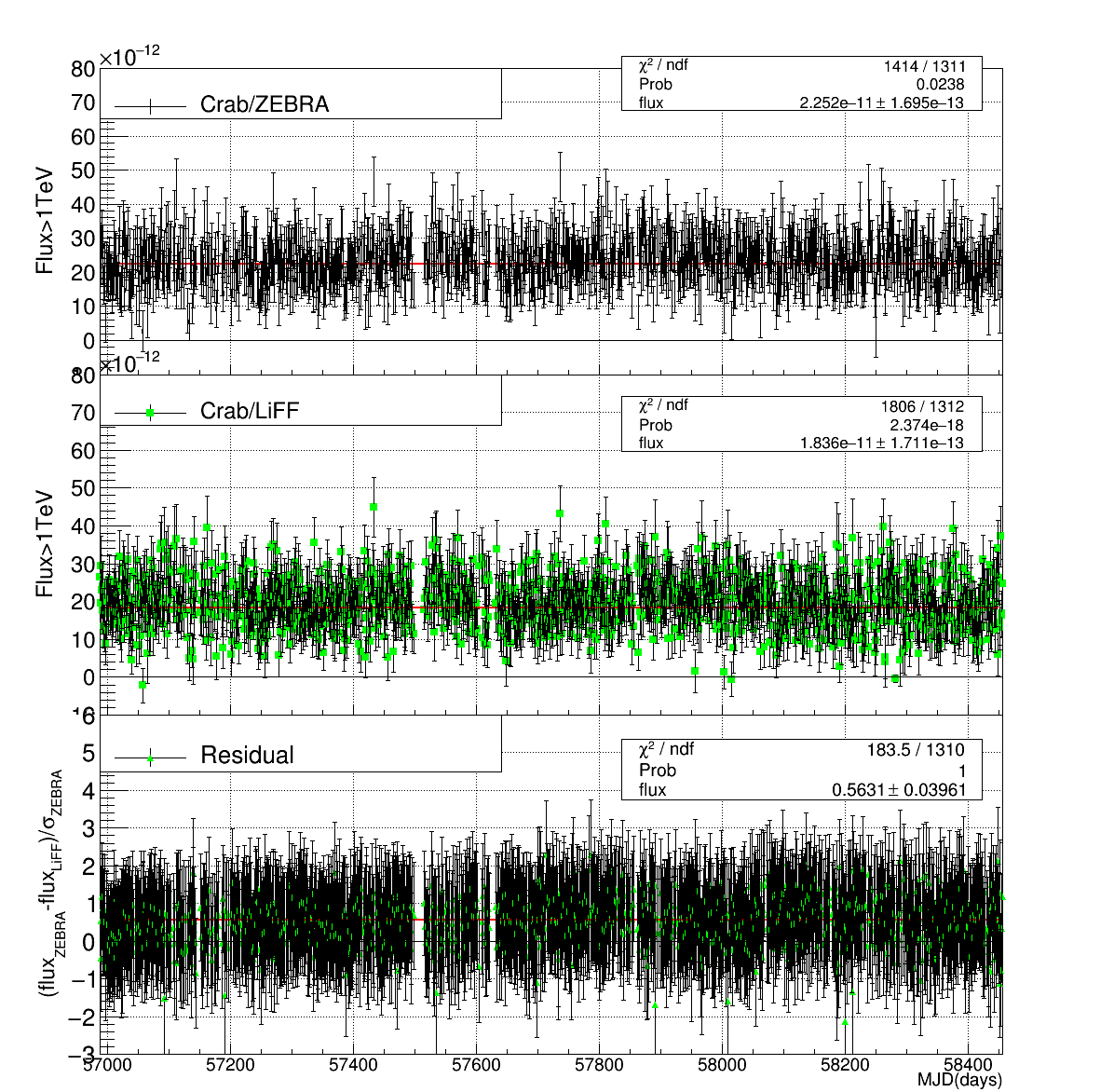}
\caption{Crab LC comparison using ZEBRA(top) and liff (middle). The difference (bottom) taken between ZEBRA and liff divided by the flux uncertainty of ZEBRA shows the consistency between both flux estimation methods.  \label{fig:compCrab}}
\end{center}
\end{figure*}

\begin{figure*}
\begin{center}
\includegraphics[width=0.75\linewidth]{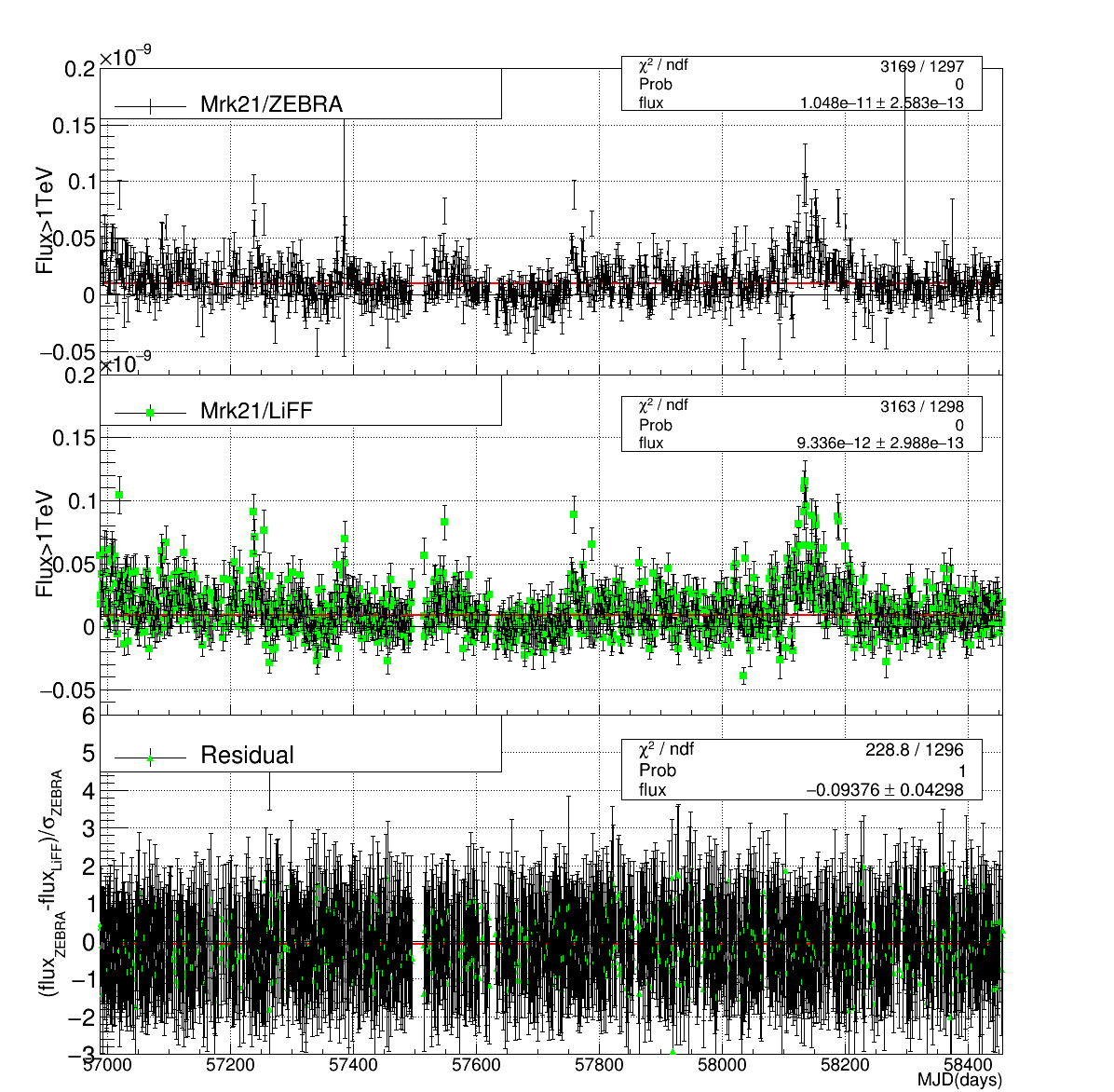}
\caption{Mrk\ 421 LC comparison using ZEBRA(top) and liff (middle). The difference (bottom) taken between ZEBRA and liff divided by the flux uncertainty of ZEBRA shows the consistency between both flux estimation methods.\label{fig:compMrk421}}
\end{center}
\end{figure*}

Figures \ref{fig:compCrab} and \ref{fig:compMrk421} show the observed LCs for the Crab and Mrk 421 respectively for the first 17 months of data used in~\cite{2017ApJ...841..100A} for ZEBRA in the top panel and LiFF in the middle panel respectively. For comparison purposes, the flux difference between results from both approaches divided over the ZEBRA uncertainty ($\rm  (flux_{zebra} - flux_{LiFF})/\sigma_{zebra}$)) for each day are shown in the lower panel of figures \ref{fig:compCrab} and \ref{fig:compMrk421}. The selection criteria and quality checks are the same used in~\cite{2017ApJ...841..100A}. For a few days we do not have flux values from both approaches as they are transits with < 0.5 coverage, and those days are not included in the lower panels. In the case of the Crab we use a simple power law with an spectral index of 2.63. For Mrk\ 421 we use a power law with a spectral index of 2.2 and an exponential cutoff at 5 eV as in~\cite{2017ApJ...841..100A}.

As observed in figure \ref{fig:compCrab}, in overall, there is a very good agreement between both LCs. There is an overall flux difference less than \%10 that is consistent with the systematic uncertainties considered for LiFF in the 17 months analysis~\cite{2017ApJ...841..100A}. The flux obtained by ZEBRA for the Crab is consistent with the values reported by High Energy Stereoscopic System (HESS) $>1$~TeV~\cite{2006A&A...457..899A}


Some features can be observed: i) These LCs show consistency with high and low flux states when comparing both methods on the same source, ii) The flux computed for either method for Crab and Mrk 421 do not exhibit variations larger than 2 $\sigma$ and iii) Considering their uncertainties, both methods on each source show LCs  that have similar average fluxes.


%
%
%
%

\section{Conclusions\label{sec:conc}}
We have proved that both methods LiFF and ZEBRA give consistent results for a flux measurement. We have reproduced the LCs presented in~\cite{2017ApJ...841..100A} for the Crab and Mrk\ 421 using ZEBRA and any future analysis that involved these sources will benefit of this new implementation of ZEBRA. Since we have consistent LCs we can be confident that the physics derived from these results will be the same as the one obtained in~\cite{2017ApJ...841..100A} therefore, Bayesian Blocks analysis among some other important measurements will be carried out using the new fitting framework such as X-ray/$\gamma$-ray correlation and variability studies. 


\begin{large}
\section*{Acknowledgements}
We acknowledge the support from: the US National Science Foundation (NSF); the US Department of Energy Office of High-Energy Physics; the Laboratory Directed Research and Development (LDRD) program of Los Alamos National Laboratory; Consejo Nacional de Ciencia y Tecnolog\'ia (CONACyT), M\'exico (grants 271051, 232656, 260378, 179588, 254964, 271737, 258865, 243290, 132197, 281653)(C\'atedras 873, 1563, 341), Laboratorio Nacional HAWC de rayos gamma; L'OREAL Fellowship for Women in Science 2014; Red HAWC, M\'exico; DGAPA-UNAM (grants AG100317, IN111315, IN111716-3, IA102715, IN111419, IA102019, IN112218), VIEP-BUAP; PIFI 2012, 2013, PROFOCIE 2014, 2015; the University of Wisconsin Alumni Research Foundation; the Institute of Geophysics, Planetary Physics, and Signatures at Los Alamos National Laboratory; Polish Science Centre grant DEC-2014/13/B/ST9/945, DEC-2017/27/-\

\noindent B/ST9/02272; Coordinaci\'on de la Investigaci\'on Cient\'ifica de la Universidad Michoacana; Royal Society-Newton Advanced Fellowship 180385. 
Thanks to Scott Delay, Luciano D\'iaz and Eduardo Murrieta for technical support.
\end{large}

\begin{large}
\addcontentsline{toc}{chapter}{References}

\bibliographystyle{plain}

\end{large}
\end{document}